\documentclass[usenatbib,usegraphicx]{mn2e}
\def\lsim{\raise0.3ex\hbox{$<$\kern-0.75em\raise-1.1ex\hbox{$\sim$}}}
\def\gsim{\raise0.3ex\hbox{$>$\kern-0.75em\raise-1.1ex\hbox{$\sim$}}}

\title{The Earliest Epoch of Reionisation in the Standard
$\Lambda$CDM Model}
\author[M. Fukugita and M. Kawasaki]
            {M. Fukugita$^{1}$ and M. Kawasaki$^2$\\
             $^1$Institute for Cosmic Ray Research, University of Tokyo,
               Kashiwa 277-8582, Japan\\
            $^2$Research Center for the Early Universe,
            Graduate School of Science, University of Tokyo,
            Tokyo 113-0033, Japan}
\date{Accepted  Received }
\pagerange{\pageref{firstpage}--\pageref{lastpage}}
\pubyear{2003}

\begin{document}

\label{firstpage}

\maketitle

\begin{abstract}
We show that the earliest possible reionisation of the Universe is
approximately at $z\simeq 13.5$ and the optical depth is $\tau\simeq
0.17$
in the conventionally accepted $\Lambda$
cold dark matter ($\Lambda$CDM) model with adiabatic fluctuations of
a flat spectrum normalised to the cosmic microwave background.
This is consistent with the reionisation found by WMAP
(the apparently earlier reionisation epoch of the WMAP is ascribed to
the adoption of the instantaneous reionisation approximation), i.e.,
the WMAP result is realised only if reionisation of the universe takes place
nearly at the maximal efficiency in the $\Lambda$CDM model.

\end{abstract}


\section{Introduction}

The most remarkable discovery from the {\it Wilkinson Microwave
Anisotropy Probe} (WMAP) Observation is a very early reionisation
of the universe at $z\simeq 20$  (Bennett et al. 2003).
The optical depth is $\tau=0.17\pm0.04$ (Kogut et al. 2003) indicated by
the TE cross power correlation function for low $\ell$ modes,
or $\tau=0.17{+0.08\atop-0.07}$ from a global
fit including both TE and TT power spectra (Spergel et al. 2003).

It is now generally accepted that early reionisation is primarily due
to UV light from early OB stars in protogalaxies (Couchman \& Rees 1986;
Tegmark et al. 1994; Cen \& Ostriker 1992;
Fukugita \& Kawasaki 1994; see Loeb \& Barkana
2001 for reviews of the recent progress and further references).
Many calculations have been done with increasingly improved
physical approximations (see Cen 2003; Razoumov et al. 2002;
Ciardi, Ferrara \& White 2003 for the most recent calculations).
Recent calculations are mostly based
on $N$-body simulations to include many physical effects.
With the cosmological parameters of $\Lambda$CDM models,
the reionisation epoch is inferred to be $z\sim 8-11$.
Although these estimates stand for the `best estimates' with elaborated
simulations, there are some subtleties that may depend on specific
assumptions, models of physical processes and meshes of numerical
simulations.
In this paper we study the {\it earliest} possible reionisation epoch
in the $\Lambda$CDM scenario under the standard assumptions concerning
star formation. Namely, we consider the case when all energy from
baryons
that form stars are efficiently injected into the intergalactic medium.
For this purpose it is appropriate to take a homogeneous multiphase
universe model, where we reduce the detailed physics to a few
parameters,
which are constrained by empirical knowledge, or otherwise left as
free parameters.

The cosmological parameters are now well-constrained
by virtue of WMAP itself, together with advancement in optical
observations: $H_0=73$ km s$^{-1}$Mpc$^{-1}$, $\Omega_m=0.25$,
$\Omega_\Lambda=0.75$ (i.e., the flat universe), $\Omega_b=0.043$
(Spergel et al. 2003). The fluctuations show the spectrum close to
flat, $n=0.97\pm0.03$, with the amplitude normalisation
$\sigma_8=0.8 - 0.9$. It seems
that there is no much freedom to modify these parameters.
The formation of structure is well described by
the Press Schechter formalism (Press \& Schechter 1974).
We assume that a constant fraction of baryons form stars once 
the perturbation  is
collapsed and the Jeans and cooling conditions are satisfied.
The maximally
allowed baryon fraction that goes to stars
is constrained by empirical metallicity of the universe. We assume
the Salpeter initial mass function (IMF), which
shows a rather high weight for high mass stars, relative to Scalo's IMF
(Scalo 1986), although this choice is not essential to us under our
empirical constraint from metallicity.
We allow the fraction of UV that comes out of OB stars in protogalaxies
to vary: the extreme case assumes a 100\% escaping fraction.

There are many effects that should be taken into account for
more realistic reionisation history: inhomogeneity
of star forming clouds, feedback from stars and expanding UV shell
are obvious examples.
The absorption due to galaxies and Lyman alpha clouds are another.
These effects all contribute to delay the reionisation epoch,
so that the neglect of these effects are justified for the purpose
to estimate the earliest reionisation epoch.
We discuss, however, these physical effects later.

We find that quasars are important only in the late epoch.
So we switch off quasars for our calculation of the
reionisation epoch.

We follow the formalism given in Fukugita \& Kawasaki
(1994; hereafter FK).
We solve the evolution equation for thermal history with
UV from collapsed objects taken into account. We include
the effect of heating to the Jeans mass, although it is not
very important. A full description of the basic formalism is found
in FK.
We give a brief description of our calculation and inputs that
differ from FK in the next section.

\section{Calculation}

The number of collapsed objects within the mass range $M$ to $M+dM$ at
redshift $z$,
$N(z,M)$, is calculated by the Press Schechter formula with
the threshold mass density $\delta_c=1.68$. We take a flat ($n=1$)
spectrum and the transfer function given by Bardeen et al. (1986)
including baryons.
We take the high normalisation $\sigma_8=0.9$,
allowed by the WMAP result
(Spergel et al. 2003), but later show
how the result depends on $\sigma_8$.
The smearing is made with a top hat window function.
We take cosmological parameters from the fit of WMAPext+2dGRS (with
constant power spectrum) in Spergel et al. (2003).

We assume that a constant fraction $f$ of baryons collapsed into
bound objects form stars once the Jeans condition
\begin{equation}
       M > M_{J}=1.4\times 10^{5} M_\odot \Omega_m^{-1/2}h^{-1}
       \left({ T_e \over \mu T_\gamma}\right)^{3/2},
\end{equation}
where $T_e$ is the electron temperature, $T_\gamma$ is the temperature
of the cosmic background radiation and $\mu$ is the mean molecular
weight, is satisfied, and the molecular cooling time is shorter than the
dynamical time (Blumenthal et al. 1984), i.e.,
\begin{eqnarray}
        M > M_{\rm mc} & = & 3.7\times 10^5 (\Omega_mh^2)^{-0.917}
        (Y_e/10^{-4})^{-0.625} \nonumber \\
        & & \times (\Omega_b/\Omega_m)^{-2.04}
        \left({1+z \over 10}\right)^{-2.75}M_\odot,
\end{eqnarray}
where $Y_e=10^{-5}\Omega_b^{-1}\Omega_{m}^{1/2}h^{-1}$ is the fraction
of
free electrons.
H$_2$ molecules are fragile, and this condition may have to be replaced
with the atomic cooling condition $T_{\rm vir}>10^4$K
($T_{\rm vir}$: virial temperature) for a more realistic calculation
(Stecher \& Williams 1967; Haiman et al. 1997). The atomic cooling
condition gives the lower bound on the galactic mass as
\begin{equation}
       M > M_{\rm ac} = 2.1\times 10^9M_{\odot} (\Omega_mh^2)^{-1/2}
       (1+z)^{-3/2},
\end{equation}
but the results using this condition differ very little from those
we obtained by setting the
molecular cooling condition.
We take the object with ${\rm max}[M_{\rm mc}, M_J]<M$ 
or ${\rm max}[M_{\rm ac}, M_J]<M$ as being collapsed.
The cooling time becomes longer than the Hubble time for low redshift
$1+z< 7 (\Omega_mh^2)^{1/5}$, where Compton cooling is not efficient,
and galaxies do not form if
\begin{equation}
       M>3.0\times 10^{14}M_\odot (1+z)^{3/4}(\Omega_b h^2)^{3/2}
       (\Omega_m h^2)^{-5/4}(Z/0.01)^{3/2},
\end{equation}
$Z$ being the metallicity (Blumenthal et al. 1984).
The constant fraction $f$ may be estimated by the balance of
the local infall rate and the cooling rate, which, however,
may depend on the details of calculations
(e.g., Cen \& Ostriker 1992).
Here we simply constrain the $f$ parameter from the heavy
element abundance to avoid uncertainties from the use
of specific models. [Note that our $f$ is the fraction of
baryons that form stars against collapsed baryons that satisfy
cooling conditions, in contrast to the definition often taken
in the literature that a similar number is defined by the ratio of
baryons that satisfy the cooling conditions (which equal baryons that
form stars) against those in collapsed objects.]

We assume that stars form according to
the Salpeter IMF, $\phi(M_s)$.
The temperature-mass relation (Bond, Carr \& Hogan 1986)
we adopt is
\begin{equation}
       T_s(M_s)=6\times 10^4{\rm K}~{\rm min}
       \left[\left({M_s\over100M_\odot}\right)^{0.3}, 1\right]
\label{eq:temp}
\end{equation}
for population II stars ($Z < 0.001$). The coefficient is replaced with
$4.3\times 10^4$K for population I stars ($Z>0.01$). For $0.001<Z<0.01$
we consider a mixture of the two populations to adjust the mean 
metalliciy.
In principle, somewhat higher temperature
is expected for so-called population III stars, i.e.,
stars from pristine gas, but the local environment of star forming 
region
is quickly enriched once the first stars formed:
so it is more appropriate to consider
population II stars here\footnote{If we take account of population III
stars for $Z<10^{-4}$, assuming that star forming regions are not
locally enriched after the first stars formed, the reionisation
epoch is only slightly delayed ($z_{\rm reion}= 13$, $\tau=0.16$, 
which are comapred to the numbers given in section 3 below).
This results from the fact that the temperature of population III stars
is higher (the coefficient of (\ref{eq:temp}) being 10), but the number 
of photons is fewer 
(Carr, Bond \& Arnett 1984), and these two effects
counteract for reionisation (the latter effect slightly wins);
the efficiency of reionisation
is more importantly controlled by the energy generation rate, 
which is fixed by
(\ref{eq:eps-t}).}.
The fraction of radiation energy produced,
$\epsilon_s M_s$, and the time
in the main-sequence, $t_{\rm MS}$, are, respectively,
(Bond, Carr \& Hogan 1986)
\begin{eqnarray}
        \epsilon_s &=& 0.0046
        \left(\frac{X}{0.76}\right)~{\rm min}
        \left[\left({M_s\over 100M_\odot}\right)^{1/2},1\right], \cr
        t_{\rm MS} &=& 2.3\times 10^6 {\rm yr}
        \left({\epsilon_s\over0.0046} \right)
        {\rm max}\left[\left({M_s\over 100M_\odot}\right)^{-2},1\right],
\label{eq:eps-t}
\end{eqnarray}
where $X=0.76$ is the hydrogen mass fraction.
Since the lifetime of massive main-sequence stars with $M_s>10M_\odot$,
which are relevant to us, is shorter than the cosmic time for
$z<50$, we can assume that ionising photons are produced instantaneously
upon the formation of stars. The production rate of UV photons with
energy $\varepsilon_\gamma$ is given by
\begin{eqnarray}
       \left[{dn_\gamma(\varepsilon_\gamma)\over dz}\right]_{r, star}
        & = &
       \int dM_s {B(\varepsilon_\gamma, T_s)\over \varepsilon_\gamma}
       \epsilon_s \phi(M_s){\Omega_b \over \Omega_m}
       \nonumber\\
       &  & \times \int dM M
       \left[{\partial N(M,z)\over \partial z}\right].
       \label{eq:ngamma}
\end{eqnarray}
where $B(\varepsilon_\gamma,T_s)$ is the blackbody spectrum,
normalised to $\int d\varepsilon_\gamma B(\varepsilon_\gamma,T_s)=1$.
For simplicity of calculation we use the derivative of the
Press-Schechter function $N$, whilst being aware that it does not give
the formation rate of collapsing objects when the destruction rate
gives a negative contribution to the derivative of $N$.
At early epochs, $z>5$, however, this never happens for galactic scales
or smaller scales (Kitayama \& Suto 1996).
Thus, this does not introduce
appreciable inaccuracies in the calculation of early reionisation.

We follow the chemical evolution of the universe.
Stars with $M_s>4M_\odot$ eject
heavy elements at a fraction
\begin{eqnarray}
       Z_{\rm ej} &=& 0.5-(M_s/6.3M_\odot)^{-1} ~~ {\rm for}~
          15<M_s/M_\odot<100\cr
       Z_{\rm ej} &=& 0.1~~~~~~~~~~{\rm for}~
          8<M_s/M_\odot<15 \cr
       Z_{\rm ej} &=& 0.2~~~~~~~~~~{\rm for}~
          4<M_s/M_\odot<8 ,
\end{eqnarray}
(Carr, Bond \& Arnett 1984).
The evolution of heavy element is then
\begin{equation}
       {dZ\over dz}=\int dM_s Z_{\rm ej} \phi(M_s) f
       {\Omega_b\over \Omega_m}
       \int dM M
       \left[{\partial N(M,z)\over \partial z}\right].
       \label{eq:metal}
\end{equation}
We take $Z\leq 0.02$ as the limit for population I stars,
which leads to $f\leq 0.55$.

We then solve the coupled evolution equations for ionisation fraction
of hydrogen and helium,
$d[n_{\rm HII}/n_{\rm H}]/dt$, $d[n_{\rm HeII}/n_{\rm He}]/dt$,
$d[n_{\rm HeIII}/n_{\rm He}]/dt$,
the photon spectrum $dn_\gamma(\varepsilon_\gamma)/dt$ and
the electron temperature $d[kT_en_b/\mu]/dt$. We take all
relevant atomic processes for heating and cooling.
The explicit
form of differential equations and necessary atomic physics
are detailed in FK.
We follow the thermal history from $z > 1000$ including the
recombination epoch.

\section{Results and Discussion}

The results of calculations are shown in Figure 1 (for hydrogen) and in
Figure 2 (for helium). The figure shows reionisation of hydrogen
and excitation of helium I to II at $z_{\rm reion}=13.5$, taking 90\%
ionisation
as the reionisation epoch.
We define the Thomson optical depth by
\begin{equation}
       \tau=\int^\infty_0 dz c \sigma_T \left(dt\over dz\right)
       \left[n_{\rm HII}-n_{\rm HII}({\rm no~reionisation})\right],
\label{eq:thomson}
\end{equation}
with $\sigma_T$ the Thomson cross section. Here, we subtract the
contribution to $\tau$ from residual ionisation of the standard
recombination (the second term in the integrand) to evaluate
only the effect of reionisation. We obtain
$\tau=0.172$. We consider that these $z_{\rm reion}$
and $\tau$ correspond to the
earliest epoch of reionisation and the maximum
optical depth we can obtain in the $\Lambda$CDM model with the specified
cosmological parameters and
with flat adiabatic fluctuations normalised to cosmic microwave
background  observations.
The optical depth is consistent with the WMAP result $\tau=0.17\pm 
0.04$,
but the reionisation epoch is later than $z_{\rm reion}=17$
the WMAP team quoted,
which assums instantaneous reionisation (Kogut et al. 2003).
[We note that $z_{\rm reion}\approx 20$ often quoted by
the WMAP team (Bennett et al. 2003; Kogut et al. 2003) assumes
a rather unusual reionisation history.]
The difference in $z_{\rm reion}$ comes from the fact
that stars formed earlier
than the reionisation epoch contribute to $\tau$. Although the HI 
fraction
drops quite sharply at $z_{\rm reion}$, there is a
non-negligible contribution from early
formation of stars at $z>z_{\rm reion}$.
If we integrate (\ref{eq:thomson}) to z=13.5, we
obtain $\tau=0.13$, i.e. $\Delta\tau=0.04$ comes from $z>13.5$.

\begin{figure}
          \centering
          \includegraphics[width=8.5cm]{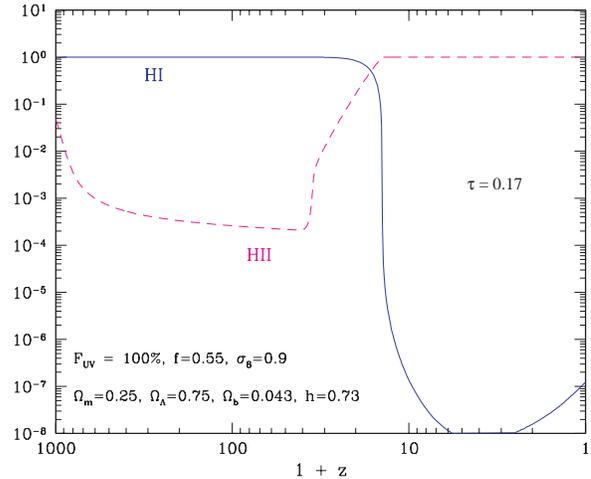}
          \caption{Ionisation history for hydrogen.
          The solid (dashed) curve
          represents number fraction of HI (HII).}
          \label{fig:Hydrogen}
\end{figure}

\begin{figure}
          \vspace*{-1cm}
          \centering
          \includegraphics[width=8.5cm]{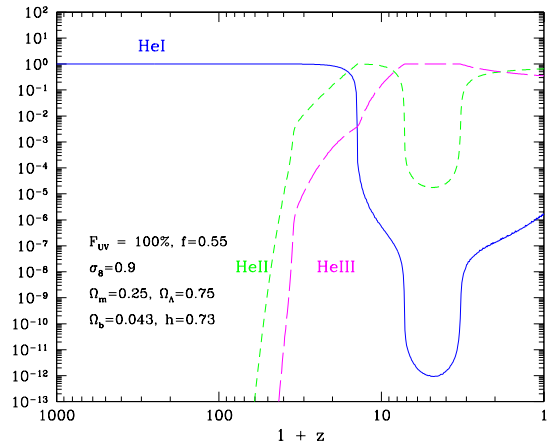}
          \caption{Ionisation history for helium. The solid, dashed
          and long-dashed curves represent number fractions of
          HeI, HeII and HeIII,
          respectively}
          \label{fig:Helium}
\end{figure}

Note that first stars formed at $z\simeq 35$, which is visible
as a turn on of the HII component. This epoch is  consistent
among the authors who calculated reionisation with the CDM model
(see e.g., Loeb \& Barkana 2001).

In our calculation we have ignored many physical effects,
which always
contribute to delay the reionisation. We have assumed as a default
that all
UV photons escape from galaxies.  This is not realistic, but this
fraction
is poorly known observationally, ranging from $>$10\% (Steidel et al.
2001)
to $<2$\% (Heckman et al. 2001).
Bianchi et al. (2001) argued $<20$\% from the Lyman $\alpha$ cloud
abundance.
This parameter may also depend on the size, and hence redshift, of
objects
we consider. We take this as a free parameter.
If we assume an escaping fraction of ionising photons
$F_{UV}=10$\% the epoch
of reionisation is delayed
to $z_{\rm reion}=11$ and $\tau=0.13$, as we shall show below.
We have assumed that stars form when the molecular
cooling condition is
satisfied. If the atomic cooling condition gives a more important
epoch for first galaxies (Haiman et al. 1997),
the epoch of reionisation is delayed,
but our calculation shows
it is only by $\Delta z\simeq 0.006$, or $\Delta\tau=-0.005$.

We have assumed the Salpeter IMF which gives a relatively
large number of massive
stars. The specific choice of this IMF, however, results in little
uncertainty, because we have set a parameter that controls star
formation
efficiency $f$ so that it gives empirically observed metallicity.
Since dominant
metal production is due to massive stars, the change caused by a
different choice of IMF, say Scalo's or top-heavy IMF, is compensated
by the change of the $f$ parameter.

We note that our results depend not only on $f$ and
$F_{\rm UV}$, but also on other cosmological parameters,
$\Omega_m$ and $\Omega_b$, and in particular on $\sigma_8$.
To show sensitivity to various parameters we show in
Figure 3 the dependences of the reionisation epoch and the optical
depth upon those parameters. The dependence on $f$ is identical to that
on $F_{UV}$.
The dependence of the reionisation epoch
on $\Omega_m$, for example,
arises from the fact that it controls the formation of small
objects in early
epochs, but on the other hand the decrease of the time interval for a 
unit
redshift as increasing $\Omega_M$ suppresses the increase of $\tau$.
The dependence on $\Omega_b$ appears
more non-trivial:
as baryon density increases, recombination rate becomes faster, delaying
reionisation. However, this effect is overcome by the increase
of the baryon density, and $\tau$ increases.
One may use this figure to compare our calculation
with those that use different cosmological parameters.

\begin{figure}
          \centering
          \includegraphics[width=8.5cm]{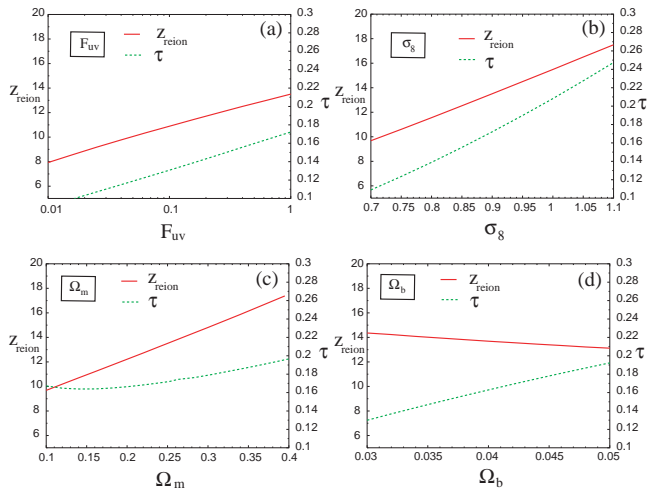}
          \caption{Dependences of the reionisation epoch
          (indicated by solid curves) and the Thomson optical
          depth (dotted curves) on the parameters of our model, (a)  
$F_{UV}$
          (b)  $\sigma_8$, (c) $\Omega_m$ and (d) $\Omega_b$.}
          \label{fig:dependence}
\end{figure}

The other effects that must be taken into account for more realistic
calculations are feedback from stars
to suppress star formation and the absorption of ionising flux by
hydrogen clouds and the envelope of galaxies. The former leads to
a milder slope of the luminosity function compared with the dark
matter mass function (Dekel \& Silk 1986; White \& Frenk 1991;
Nagamine et al. 2001).
We studied the effect by artificially decreasing the net
star formation rate for smaller objects so that the collapsed
baryonic mass function
obeys the form of an empirical Schechter function at $z=0$,
assuming $M/L=$constant.
This suppresses the early star formation
and the ionisation does not take place until $z_{\rm reion}\sim 11$
and $\tau\sim 0.12$.
This is probably an extreme case because a sharper faint end slope is 
expected
in the luminosity function at a higher $z$.

The UV absorption by Lyman alpha clouds and galaxies may delay
reionisation. We developed a formalism to take account of envelope
of galaxies and Lyman $\alpha$ clouds (Fukugita \& Kawasaki, in
preparation).
For an early epoch such as the one we are concerned with, this effect is negligible. The
absorption due to galaxies becomes sizable only at low redshifts ($z
\lsim
5$).
Adopting a minihalo model for Lyman $\alpha$ clouds
(Ikeuchi 1986; Rees 1986),
we find that absorption by Lyman $\alpha$ cloud is negligible
for the ionisation epoch, because the onset of
the ionising radiation decreases rapidly the abundance
of Lyman $\alpha$ clouds.

The ionisation by quasars is important only for $z<3$. Recent 
Sloan Digital Sky Survey (SDSS)
observations showed that early bright quasars form exponentially in
$z$ (Fan et al. 2001). We estimate the early quasar population using
the quasar luminosity function (Boyle et al. 2000)
known at lower redshifts,
and put $[dn_\gamma/dz]_{\rm qso}$ with the aid of the empirical
spectrum (Vanden Berk et al. 2001) in the evolution equation. We find
that quasars alone reionise the universe only at $z=2.6$ although
the effect becomes recognisable at $z\approx 10$. In any case the
contribution of quasars is negligible compared to OB stars
in the early reionisation epoch. The UV flux from quasars is
important to sustain a high degree of ionisation at lower redshifts,
as envisaged by the Gunn-Peterson test, when the escaping fraction
of UV from galaxies is less than $10$\%.

In conclusion, the earliest epoch of reionisation we obtained is
$z=13.5$ with the
cosmological parameters determined by WMAP (a value of 14 is
possible within
errors -- for example with $\Omega_m=0.3$) 
and the Thomson optical depth is 0.17.
This is consistent with the reionisation WMAP found. The 
difference in
$z$ beweeen our value and WMAP's is ascribed to different 
approximations:
WMAP assumed that the entire optical depth comes from $z<z_{\rm reion}$,
whereas the contribution earlier than $z=z_{\rm reion}$ is 
non-negligible
in our calculation. The realistic UV escaping
fraction and the feedback of stars may somewhat lower the optical depth
and delay the recombination epoch. The uncertainty in IMF that directly
controls star formation is effectively absorbed into the $f$ parameter
which is constrained by metallicity.
We remark that reionisation takes place sharply once the number of
ionising photons, which are determined by the thermal balance, reaches
some critical number and so it is difficult to sustain a partial 
ionisation
for an extended period, as was assumed in Kogut et al. (2003).

As the final comment our result is consistent
with the Salpeter IMF case of Ciardi et al. (2003), who carried
out a numerical simulation for the purpose similar to ours.
As a check we calculated the case with $\Omega_m=0.3$,
$\sigma_8=0.9$ and $F_{UV}=0.2$, and we obtained the
reionisation epoch of $z=12.8$, which is compared to $\simeq 12$
(see Fig.~3 of Ciardi et al.).
This small difference is probably ascribed to feedback
effects of stars, as discussed above.

\vspace{6mm}
\noindent
{\bf ACKNOWLEDGMENT}

\vspace{3mm}
\noindent
MF wishes to thank Richard Ellis and Astronomy Group of Caltech
for their kind hospitality while this work was done.


\begin{thebibliography}{999}
\bibitem[\protect\citeauthoryear{Bardeen et al.}{1986}]{Bardeen}
       Bardeen, J. M., Bond, J. R., Kaiser, N., Szalay, A. S. 1986,
       ApJ, 304, 15.
\bibitem[\protect\citeauthoryear{Bennet et al.}{1999}]{Bennet}
       Bennett, C. L. et al. 2003, astro-ph/0302207
\bibitem[\protect\citeauthoryear{Bianchi et al.}{2001}]{Bianchi}
       Bianchi, S., Christiani, S. Kim, T.-S. 2001, A\&A, 376, 1
\bibitem[\protect\citeauthoryear{Blumenthal et al.}{1984}]
       {Blumenthal}
       Blumenthal, G. R., Faber, S. M., Primack, J. R., Rees, M. J., 
1984,
       Nat, 311, 517
\bibitem[\protect\citeauthoryear{Bond, Car \& Hogan}{1986}] {Bond}
       Bond, J.R., Carr, B.J., Hogan, C.J. 1986 ApJ, 306, 428
\bibitem[\protect\citeauthoryear{Boyle et al.}{2000}] {Boyle}
       Boyle, B. J., Shanks, T., Croom, S. M., Smith, R. J., Miller, L.,
       Loaring, N., Heymans, C. 2000, MNRAS, 317, 1014
\bibitem[\protect\citeauthoryear{Car, Bond \& Arnett}{1984}] {Carr}
        Carr, B.J., Bond, J.R., Arnett, W.D., 1984 ApJ, 277, 445
\bibitem[\protect\citeauthoryear{Cen}{2003}]{Cen2}
       Cen, R. 2003, ApJ, 591,12
\bibitem[\protect\citeauthoryear{Cen}{1992}]{Cen}
       Cen, R., Ostriker, J. P. 1992, ApJ, 393, 22
\bibitem[\protect\citeauthoryear{Ciardi, Ferrara \& White}{2003}]
       {Ciardi}
       Ciardi, B., Ferrara, A., White, S. D. M., astro-ph/0302451
\bibitem[\protect\citeauthoryear{Couchman \& Rees}{1986}]
       {Couchman}
       Couchman, H. M. P., Rees, M. J. 1986, MNRAS, 221, 53
\bibitem[\protect\citeauthoryear{Dekel \& Silk}{2000}]{Haarsma}
       Dekel, A., Silk, J. 1986, ApJ, 303, 39
\bibitem[\protect\citeauthoryear{Fan et al.}{2001}]
       {Fan}
       Fan, X. et al. 2001, AJ, 121, 54
\bibitem[\protect\citeauthoryear{Fukugita \& Kawasaki}{1994}]{FK}
       Fukugita, M., Kawasaki, M. 1994, MNRAS, 269, 563
\bibitem[\protect\citeauthoryear{Haiman et al.}{1997}]{Cardelli}
       Haiman, Z., Rees, M. J., Loeb, A. 1997, ApJ, 476, 458
\bibitem[\protect\citeauthoryear{Heckman et al.}{2001}]{Heckman}
       Heckman, T. M., Sembach, K. R., Meurer, G. R., Leitherer,C.,
       Calzetti, D., Martin, C. L., 2001. ApJ, 558, 56
\bibitem[\protect\citeauthoryear{Ikeuch}{1986}]{Ikeuchi}
       Ikeuchi, S.  1986, Ap\&SS, 118, 509
\bibitem[\protect\citeauthoryear{Kitayama \& Suto}{1996}]{Kitayama}
       Kitayama, T., Suto, Y. 1996, MNRAS, 280, 638
\bibitem[\protect\citeauthoryear{Kogut et al.}{2003}]{Kogut}
       Kogut, A. et al. 2003, astro-ph/0302213
\bibitem[\protect\citeauthoryear{Loeb \& Barkana}{2002}]{Loeb}
       Loeb, A., Barkana, R. 2001, ARA\&A, 39, 19
\bibitem[\protect\citeauthoryear{Nagamine et al.}{2001}]{Hartmann}
       Nagamine, K., Fukugita, M., Cen, R.,  Ostriker, J. P. 2001, MNRAS,
       327, L10
\bibitem[\protect\citeauthoryear{Press \& Schechter}{1974}]{PS}
       Press, W. H., Schechter, P. 1974, ApJ, 187, 425
\bibitem[\protect\citeauthoryear{Razoumov et al.}{2002}]{Razoumov}
       Razoumov, A. O., Norman, M. L., Abel, T., Scott, D. 2002,
       ApJ, 572, 695
\bibitem[\protect\citeauthoryear{Rees}{1986}]{Rees}
       Rees, M. 1986, MNRAS, 218, 25p
\bibitem[\protect\citeauthoryear{Scalo}{1986}]{Scalo}
       Scalo, J. 1986, Fund. Cosmic Phys., 11, 1
\bibitem[\protect\citeauthoryear{Spergel et al.}{2003}]{Spergel}
       Spergel, D. N. et al. 2003, astro-ph/0302209
\bibitem[\protect\citeauthoryear{Stecker \& Williams}{1967}]{Stecker}
       Stecher, T.P., Williams, D.A. 1967. ApJ, 149, L29
\bibitem[\protect\citeauthoryear{Steidel et al.}{2001}]{Steidel}
       Steidel, C. C., Pettini, M., Adelberger, K. L. 2001, ApJ, 546, 665
\bibitem[\protect\citeauthoryear{Tegmark et al}{1994}]{Tegmark}
       Tegmark, M., Silk, J., Blanchard, A. 1994, ApJ, 420, 484
\bibitem[\protect\citeauthoryear{Vanden Berk et al.}{1999}]{Berk}
       Vanden Berk, D. E. et al. 2001, AJ, 122, 549
\bibitem[\protect\citeauthoryear{White \& Frenk}{1991}]{White}
       White, S. D. M., Frenk, C. S. 1991, ApJ, 379, 52
\end{thebibliography}
\end{document}